\documentclass[aps,pra,onecolumn,superscriptaddress,notitlepage,showpacs,showkeys]{revtex4-1}
\usepackage{graphicx,amsmath,amsfonts,amssymb,upgreek,txfonts,color}
\usepackage[colorlinks,linkcolor=blue,citecolor=blue,urlcolor=blue,breaklinks=true]{hyperref}
\usepackage{color, colortbl}
\usepackage{booktabs}
\definecolor{lightgreen}{rgb}{0.88,1,1}

\newcommand{\be}{\begin{equation}}
\newcommand{\ee}{\end{equation}}
\newcommand{\bae}{\begin{eqnarray}} \newcommand{\eae}{\end{eqnarray}}

\begin{document}


\title{Controlling multiparameter quantum estimation in exciton-optomechanics system}

\author{Hamza Harraf}
\affiliation{LPHE-Modeling and Simulation, Faculty of Sciences, Mohammed V University in Rabat, Rabat, Morocco}	
\author{Mohamed Amazioug} 
\thanks{m.amazioug@uiz.ac.ma}
\affiliation{LPTHE-Department of Physics, Faculty of sciences, Ibnou Zohr University, Agadir, Morocco}	
\author{Rachid Ahl Laamara}
\affiliation{LPHE-Modeling and Simulation, Faculty of Sciences, Mohammed V University in Rabat, Rabat, Morocco}	
\affiliation{Centre of Physics and Mathematics, CPM, Faculty of Sciences, Mohammed V University in Rabat, Rabat, Morocco}

\date{\today}

\begin{abstract}

Multiparameter quantum estimation has emerged as a central task in quantum metrology. In this work, we investigate multiparameter quantum estimation in a hybrid exciton--optomechanical (EOM) system. The system consists of a semiconductor quantum well embedded inside a driven optomechanical microcavity, where the excitonic, optical, and mechanical modes interact coherently through exciton--photon and radiation-pressure couplings. Using the Gaussian-state formalism, we derive the covariance matrix of the steady-state quantum fluctuations and employ both the symmetric logarithmic derivative (SLD) and right logarithmic derivative (RLD) approaches to evaluate the quantum Fisher information matrix associated with the simultaneous estimation of the exciton--photon coupling strength $g$ and the excitonic decay rate $k_x$. We analyze the corresponding quantum Cram\'er--Rao bounds and determine the most informative precision limit governing the attainable estimation accuracy. The influence of several experimentally relevant parameters, including temperature, driving power, optomechanical coupling strength, and dissipation rates, is investigated in detail. Our results show that strong hybrid interactions and low-temperature regimes significantly enhance the estimation precision, whereas thermal fluctuations and dissipation processes deteriorate the metrological performance. Furthermore, we compare the ultimate quantum limits with experimentally feasible Gaussian measurement strategies based on homodyne and heterodyne detection. We show that heterodyne detection provides better estimation performance than homodyne schemes and can approach the optimal quantum precision limit in suitable parameter regimes. These findings demonstrate that the interplay between coherent hybridization, dissipation, and Gaussian quantum fluctuations provides an effective mechanism for enhancing multiparameter estimation sensitivity, establishing exciton--optomechanical systems as promising platforms for quantum metrology and quantum sensing applications.

\end{abstract}

\maketitle

\section{\label{sec1}Introduction}

Quantum estimation theory has become a fundamental tool for determining the ultimate precision limits allowed by quantum mechanics~\cite{Helstrom1976,Holevo2011,Braunstein1994,Albarelli2019}. The central quantity in quantum metrology is the quantum Fisher information (QFI) \cite{Zhang2023}, which quantifies the sensitivity of a quantum state with respect to infinitesimal variations of physical parameters. The achievable estimation precision is bounded by the quantum Cram\'er-Rao bound (QCRB) \cite{Alipour2014}, establishing the minimal uncertainty attainable for a given estimation protocol~\cite{Paris2009,Giovannetti2011}. Quantum metrology has found broad applications in gravitational-wave detection, quantum thermometry, magnetometry, phase estimation, and frequency sensing~\cite{Abbott2016,Correa2015,Degen2017}. While single-parameter quantum estimation is now relatively well understood, the simultaneous estimation of multiple parameters remains considerably more challenging~\cite{Szczykulska2016,Liu2020}. In multiparameter quantum metrology, the optimal measurements associated with different parameters may become incompatible due to the noncommutativity of the corresponding quantum observables. Consequently, the simultaneous saturation of the QCRB is generally nontrivial and requires a careful analysis of the quantum Fisher information matrix (QFIM)~\cite{Ragy2016,Demkowicz2020,Arkhipov2026}. This has motivated extensive theoretical efforts aimed at identifying optimal estimation strategies capable of approaching the ultimate precision bounds in realistic physical systems. Recently, multiparameter quantum estimation has attracted increasing attention in  magnonic systems~\cite{hamza2025}. In these platforms, strong coherent interactions and controllable dissipation channels provide favorable conditions for encoding information into Gaussian quantum fluctuations. Moreover, the possibility of engineering nonclassical states through squeezed fields or nonlinear interactions offers additional opportunities for enhancing metrological precision. However, multiparameter quantum estimation in hybrid exciton-optomechanical systems remains largely unexplored. The various architectures employed in quantum metrology, continuous-variable (CV) systems constitute a particularly attractive framework owing to their experimental accessibility and analytical tractability. In this context, Gaussian states have emerged as one of the most important resources in quantum information science \cite{Ferraro2005,Weedbrook2012}. Their complete characterization through first and second statistical moments considerably simplifies the description of complex quantum systems while remaining experimentally accessible in a broad range of quantum optical and optomechanical platforms. Gaussian states have played a central role in quantum communication, quantum sensing, quantum teleportation, and cavity optomechanics \cite{Braunstein2005, Vitali2007,Paternostro2007,Aspelmeyer2014}. Furthermore, Gaussian quantum fluctuations provide an efficient carrier of parameter information and therefore constitute a natural framework for quantum metrology and multiparameter estimation.

Among the various hybrid continuous-variables platforms, exciton-optomechanical systems have emerged as highly promising candidates for investigating light-matter interactions at the quantum level~\cite{Deng2010,Restrepo2014,Vyatkin2021,Wu2024,Cai2025}. These systems combine the coherent interaction between cavity photons and excitons with radiation-pressure coupling to a mechanical oscillator, leading to rich nonlinear and quantum coherent phenomena. In particular, the strong exciton-photon interaction enables the formation of exciton-polariton modes when the coupling strength exceeds the dissipation rates of the cavity and excitonic subsystems~\cite{Kasprzak2006,Deng2002}. Such hybrid excitations provide an efficient framework for manipulating quantum fluctuations and enhancing the sensitivity of quantum optical devices. Furthermore, exciton-optomechanical systems have been widely investigated in connection with quantum entanglement generation, quantum state transfer, optical bistability, squeezing, and photon blockade effects~\cite{Tian2010,Liew2010,Rabl2011}.

Motivated by these considerations, in this work we investigate multiparameter quantum estimation in a hybrid exciton-optomechanical system. Our objective is to simultaneously estimate the exciton-photon coupling strength $g$ and the excitonic dissipation rate $k_x$ by exploiting the Gaussian quantum fluctuations of the steady state. Using the symmetric logarithmic derivative (SLD) and right logarithmic derivative (RLD) formalisms, we derive the corresponding quantum Fisher information matrices and analyze the associated quantum Cram\'er-Rao bounds. Furthermore, we examine the influence of several experimentally relevant parameters, including temperature, driving strength, dissipation rates, and optomechanical coupling, on the attainable estimation precision. We additionally compare the ultimate quantum limits with experimentally feasible Gaussian measurement strategies based on homodyne and heterodyne detection. Our results demonstrate that the hybrid exciton-optomechanical platform provides a promising architecture for high-precision multiparameter quantum metrology. In particular, we show that the interplay between coherent hybridization, dissipation, and quantum fluctuations can strongly modify the estimation sensitivity and generate regimes of enhanced precision. The present work therefore contributes to the growing effort toward the development of quantum-enhanced sensing protocols in hybrid continuous-variable systems.

This article is structured as follows. In Sec. \ref{sec2}, we analyze the system dynamics through the quantum Langevin equations (QLEs). Section \ref{sec3} is devoted to the linearization of these equations and the derivation of the covariance matrix (CM) in both the steady-state and dynamical regimes. In Sec. \ref{sec4}, we present the theoretical framework for multiparameter quantum estimation of Gaussian states, where the quantum Fisher information matrix (QFIM) is derived using the symmetric logarithmic derivative (SLD) and right logarithmic derivative (RLD) operators. We further compare individual and simultaneous estimation strategies and examine the classical Fisher information associated with homodyne and heterodyne detection schemes. The numerical results are presented and discussed in Sec. \ref{sec5}. Finally, Sec. \ref{sec7} summarizes the main findings of this work and concludes with perspectives for future research.

\section{\label{sec2} The Model}

\begin{figure}[ht!]
	\centering
\begin{tabular}{ccc}
\includegraphics[scale=0.25]{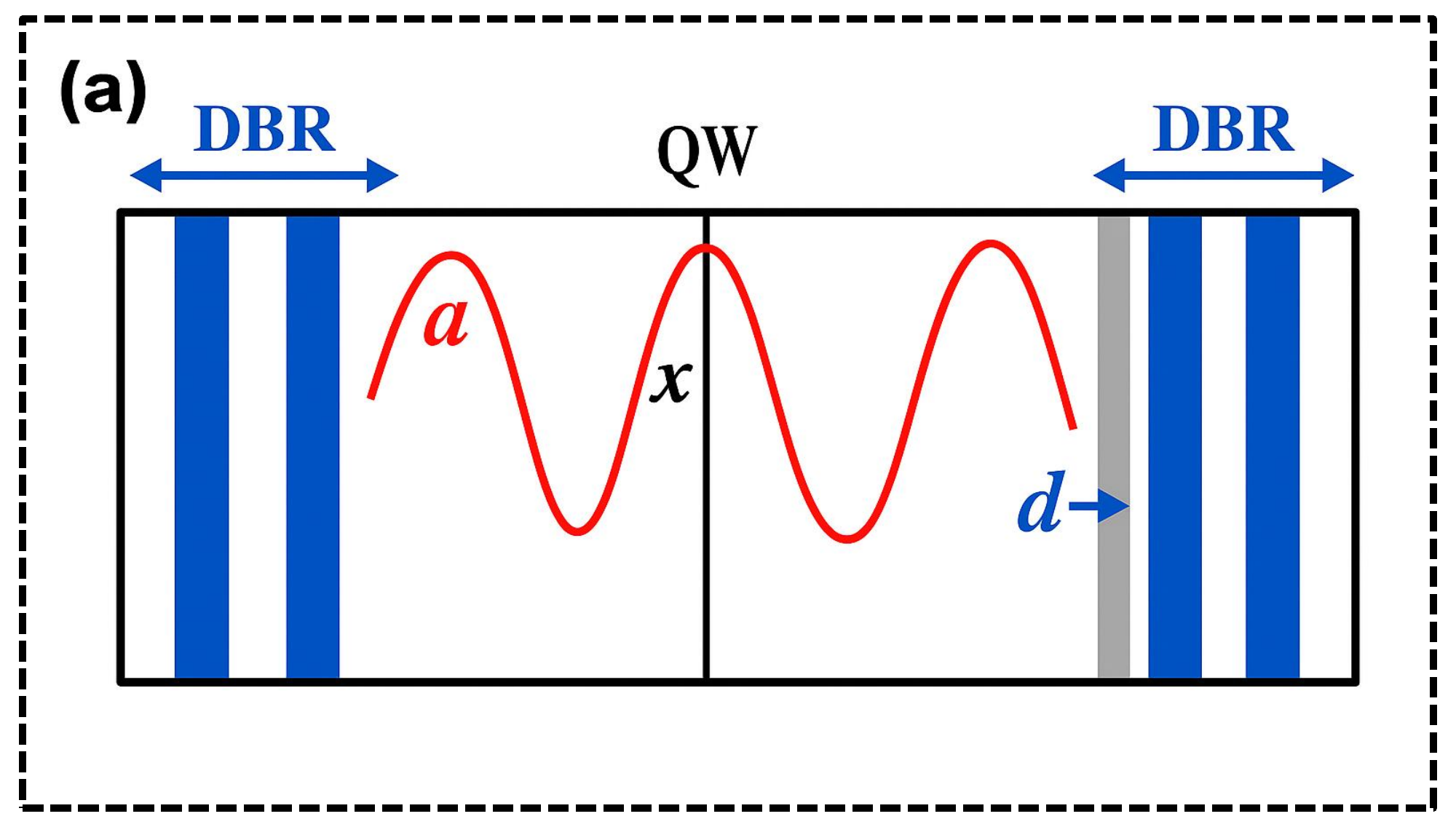}
\includegraphics[scale=0.25]{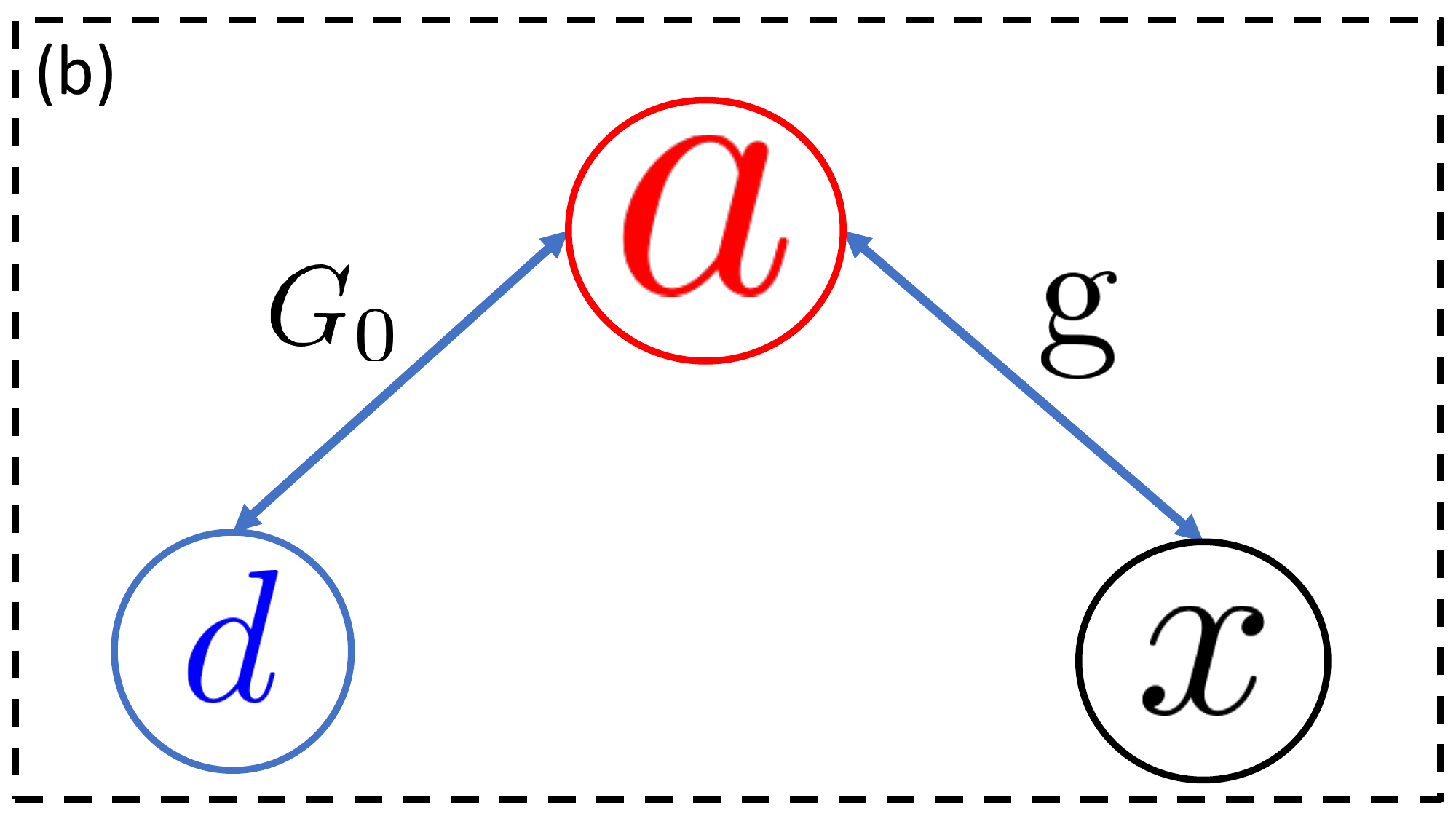}
\end{tabular}
	\caption{(a) Sketch of exciton--optomechanical (EOM). A quantum well (QW) is embedded in a semiconductor microcavity formed by two distributed Bragg reflectors (DBRs), enabling interaction between the exciton mode ($x$) and the optical cavity mode ($a$). The DBRs are movable, and their mechanical motion ($d$) couples to the cavity photons through a dispersive interaction. The shaded grey region illustrates the mechanical degree of freedom and is not intended to represent the actual physical motion. (b) The cavity mode ($a$) couples to the exciton mode ($x$) and  via a linear beam-splitter interaction ($g$). In addition, the cavity mode interact with the mechanical mode ($d$) through a dispersive electrostrictive coupling ($G_0$).}
	\label{fig1}
\end{figure}

The exciton--optomechanical (EOM) system consists of a quantum well (QW) embedded in a semiconductor microcavity formed by two movable distributed Bragg reflectors (DBRs), as illustrated in Fig.\ref{fig1}(a). A DBR is composed of alternating layers of materials with high and low refractive indices, where each layer has an optical thickness of $\lambda/4$, with $\lambda$ denoting the optical wavelength. Under this condition, the reflected light from successive interfaces undergoes constructive interference, giving rise to a photonic stop band that suppresses transmission. Consequently, the DBR behaves as a highly reflective mirror for incident light whose wavelength lies within this stop band. Since the DBRs are movable, their mechanical displacement couples dispersively to the cavity photons through the optomechanical interaction \cite{Kyriienko2014,Carlon2022,Fainstein2013}. 
The QW consists of a thin semiconductor layer confined between two barrier materials with significantly larger band gaps. When embedded inside the microcavity, the QW enables a linear excitation-exchange (beam-splitter-like) interaction between cavity photons and excitons. The Hamiltonian describing the EOM system is given by \cite{Zuo2024}

\begin{equation}
\mathcal{H}/\hbar = \Omega_x x^\dagger x + \Omega_a a^\dagger a + \Omega_d d^\dagger d + g(x^\dagger a + x a^\dagger) + G_0 a^\dagger a (d + d^\dagger) + i\varepsilon(a^\dagger e^{-i\Omega_0 t} - a e^{i\Omega_0 t}),
\end{equation}

where $x$, $a$, and $d$ ($x^\dagger$, $a^\dagger$, and $d^\dagger$) denote the annihilation (creation) operators associated with the excitonic, photonic, and phononic modes, respectively, obeying the bosonic commutation relations $[\nu,\nu^\dagger]=1$ ($\nu=x,a,d$). The corresponding resonance frequencies are represented by $\Omega_x$, $\Omega_a$, and $\Omega_d$. The parameter $g$ denotes the exciton--photon coupling strength, which depends on the position of the QW inside the microcavity and may attain the strong-coupling regime. Specifically, when the coupling strength exceeds the exciton and cavity decay rates, i.e., $g>k_x,k_a$, the system enters the strong-coupling regime, resulting in the formation of exciton polaritons \cite{Weisbuch1992,Houdre1994}.  The single-photon optomechanical coupling strength, denoted by $G_0$, is generally weak \cite{Aspelmeyer2014}. However, the effective optomechanical interaction can be substantially enhanced by coherently driving the microcavity with an intense laser field as illustrated in Fig.\ref{fig1}(b). The last term in the Hamiltonian represents the driving contribution, where $\varepsilon=\sqrt{\frac{2\mathrm{P}k_a}{\hbar\Omega_0}}$ characterizes the coupling strength between the cavity mode and the driving field of frequency $\Omega_0$ and input power $\mathrm{P}$.

By accounting for dissipation and the associated input noise affecting each mode, the dynamics of the system can be described by the following quantum Langevin equations (QLEs), written in a frame rotating at the drive frequency $\Omega_0$:
\begin{align}
\dot{x} &= -(i\Delta_x + k_x)x - iga + \sqrt{2k_x}x_{\text{in}}, \nonumber \\
\dot{a} &= -(i\Delta_a + k_a)a - igx - iG_0 a(d + d^\dagger) + \varepsilon + \sqrt{2k_a}a_{\text{in}}, \\
\dot{d} &= -(i\Omega_d + k_d)d - iG_0 a^\dagger a + \sqrt{2k_d}d_{\text{in}}, \nonumber
\end{align}

where $\Delta_x=\Omega_x-\Omega_0$ and $\Delta_a=\Omega_a-\Omega_0$ denote the exciton--drive and cavity--drive detunings, respectively, while $k_d$ represents the mechanical damping rate. The operators $\nu_{\mathrm{in}}(t)$ ($\nu=x,a,d$) correspond to the input noise associated with the excitonic, photonic, and phononic modes. These noise operators are assumed to have zero mean and satisfy the following correlation functions \cite{Gardiner2000}
\begin{align}
\langle \nu_{\mathrm{in}}(t)\nu_{\mathrm{in}}^\dagger(t') \rangle
=&
\left[N_{\nu}(\Omega_{\nu})+1\right]\delta(t-t'),\\ \nonumber
\langle \nu_{\mathrm{in}}^\dagger(t)\nu_{\mathrm{in}}(t') \rangle
=&
N_{\nu}(\Omega_{\nu})\delta(t-t'),
\end{align}
where
$
N_{\nu}(\Omega_{\nu})
=
\left[\exp\left(\frac{\hbar\Omega_{\nu}}{k_B {\rm T}}\right)-1\right]^{-1}
$ 
is the mean thermal occupation number of mode $\nu$ ($\nu=x,a,d$) in thermal equilibrium, with $\rm T$ denoting the temperature of the surrounding bath and $k_B$ being the Boltzmann constant.

Due to the strong cavity driving and the exciton--photon excitation-exchange interaction, the cavity and exciton modes acquire large coherent amplitudes, satisfying $|\langle a \rangle|, |\langle x \rangle| \gg 1$. Under these conditions, the system dynamics can be linearized around the steady-state values by decomposing each mode operator $\nu$ into its steady-state mean value and a small quantum fluctuation, namely $\nu=\langle \nu \rangle+\delta\nu$ ($\nu=x,a,d$). Neglecting second-order fluctuation terms, the QLEs can be separated into two sets of equations describing, respectively, the classical mean fields and the quantum fluctuations. Consequently, the linearized QLEs governing the quantum fluctuations take the form:

\begin{align}
\label{eq5}
\delta\dot{x} &= -(i\Delta_x + k_x)\delta x - ig\delta a + \sqrt{2k_x}x_{\text{in}}, \nonumber \\
\delta \dot{a} &= -(i\tilde{\Delta}_a + k_a)\delta a - ig\delta x - G_{ad}(\delta d + \delta d^\dagger) + \sqrt{2k_a}a_{\text{in}}, \\
\delta\dot{d} &= -(i\Omega_d + k_d)\delta d - (G_{ad}\delta a^\dagger + G_{ad}^*\delta a) + \sqrt{2k_d}d_{\text{in}}, \nonumber
\end{align}
where $\tilde{\Delta}_a = \Delta_a + 2G_0\langle d \rangle$ is the effective cavity-drive detuning including the frequency shift due to the optomechanical interaction, and $G_{ad} = iG_0\langle a \rangle$ is the effective optomechanical coupling strength. The expressions of the classical averages are given by
\begin{align}
\langle x \rangle &= \frac{-i \varepsilon g}{g^2 + (i\tilde{\Delta}_a + k_a)(i\Delta_x + k_x)}, \nonumber \\
\langle a \rangle &= \frac{\varepsilon(i\Delta_x + k_x)}{g^2 + (i\tilde{\Delta}_a + k_a)(i\Delta_x + k_x)}, \\
\langle d \rangle &= -\frac{G_0}{\Omega_d} |\langle a \rangle|^2. \nonumber
\end{align}
Using the quadrature fluctuation operators $\delta X_{\nu} = \frac{\delta \nu + \delta \nu^\dagger}{\sqrt{2}}$ and $\delta Y_{\nu} = \frac{i(\delta \nu^\dagger - \delta \nu)}{\sqrt{2}}$ ($\nu = a, x, d$), the QLEs (\ref{eq5}) can be cast into a compact matrix form
\begin{equation}
\dot{\mu}(t) = \mathcal{S} \mu(t) + \xi(t),
\end{equation}
where $\mu(t) = [\delta X_x(t), \delta Y_x(t), \delta X_a(t), \delta Y_a(t), \delta X_d(t), \delta Y_d(t)]^T$ represents the vector of quantum fluctuations,\\ $\xi(t) = [\sqrt{2k_x}X_x^{\text{in}}, \sqrt{2k_x}Y_x^{\text{in}}, \sqrt{2k_a}X_a^{\text{in}}, \sqrt{2k_a}Y_a^{\text{in}}, \sqrt{2k_d}X_d^{\text{in}}, \sqrt{2k_d}Y_d^{\text{in}}]^T$ is the vector of input noises, where $X_{\nu}^{\text{in}}$ and $Y_{\nu}^{\text{in}}$ are defined similarly as $\delta X_{\nu}$ and $\delta Y_{\nu}$ but with the noise operators $\nu_{\text{in}}$ and $\nu_{\text{in}}^\dagger$, and the drift matrix $\mathcal{S}$ is given by

\begin{equation}
\mathcal{S} = \begin{pmatrix}
-k_x & \Delta_x & 0 & g & 0 & 0 \\
-\Delta_x & -k_x & -g & 0 & 0 & 0 \\
0 & g & -k_a & \tilde{\Delta}_a & -2\text{Re} [G_{ad}] & 0 \\
-g & 0 & -\tilde{\Delta}_a & -k_a & -2\text{Im} [G_{ad}] & 0 \\
0 & 0 & 0 & 0 & -k_d & \Omega_d \\
0 & 0 & -2\text{Im} [G_{ad}] & 2\text{Re} [G_{ad}] & -\Omega_d & -k_d
\end{pmatrix}.
\end{equation}
The stability of the system is analyzed through the characteristic equation
\begin{equation}
\det(\mathcal{S}-\eta\mathbb{I})
=\eta^6+\upsilon_1\eta^5+\upsilon_2\eta^4+\upsilon_3\eta^3+\upsilon_4\eta^2+\upsilon_5\eta+\upsilon_6,
\end{equation}
which is a sixth-order polynomial due to the dimension of $\mathcal{S}$. The stability of the system can then be investigated using the Routh--Hurwitz criterion \cite{Dejesus1987}. According to this criterion, the system is stable if and only if all Hurwitz determinants are strictly positive. Therefore, the stability conditions are expressed as
\begin{align}
\upsilon_1 &> 0, \\
\upsilon_1\upsilon_2-\upsilon_3 &> 0, \\
\upsilon_1\upsilon_2\upsilon_3-\upsilon_1^2\upsilon_4-\upsilon_3^2 &> 0, \\
\;\upsilon_1^2 \upsilon_2 \upsilon_6
-\upsilon_1^2 \upsilon_4^2
-\upsilon_1 \upsilon_2^2 \upsilon_5
+\upsilon_1 \upsilon_2 \upsilon_3 \upsilon_4
-\upsilon_1 \upsilon_3 \upsilon_6
+2 \upsilon_1 \upsilon_4 \upsilon_5
+\upsilon_2 \upsilon_3 \upsilon_5
-\upsilon_3^2 \upsilon_4
-\upsilon_5^2 &> 0, \\
\Delta_5 &> 0, \\
\Delta_6 &> 0,
\end{align}
where $\Delta_5$ and $\Delta_6$ denote the higher-order Hurwitz determinants 
\[
\Delta_5=-\upsilon_1^3\upsilon_6^2
-\upsilon_5 \Big[
\upsilon_4(\upsilon_1^2 \upsilon_4 - \upsilon_1 \upsilon_2 \upsilon_3 + \upsilon_3^2)
+ \upsilon_1 \upsilon_5 (\upsilon_2^2 - 2\upsilon_4)
- \upsilon_2 \upsilon_3 \upsilon_5
+ \upsilon_5^2
\Big]
+ \upsilon_6 \Big[
\upsilon_1^2 (2\upsilon_2 \upsilon_5 + \upsilon_3 \upsilon_4)
- \upsilon_1 \upsilon_3 (\upsilon_2 \upsilon_3 + 3\upsilon_5)
+ \upsilon_3^3
\Big],
\]
\[
\Delta_6 =
\upsilon_6 \Big(
-\upsilon_1^3 \upsilon_6^2
-\upsilon_5 \big[
\upsilon_4(\upsilon_1^2 \upsilon_4 - \upsilon_1 \upsilon_2 \upsilon_3 + \upsilon_3^2)
+ \upsilon_1 \upsilon_5 (\upsilon_2^2 - 2\upsilon_4)
- \upsilon_2 \upsilon_3 \upsilon_5
+ \upsilon_5^2
\big]
+ \upsilon_6 \big[
\upsilon_1^2 (2\upsilon_2 \upsilon_5 + \upsilon_3 \upsilon_4)
- \upsilon_1 \upsilon_3 (\upsilon_2 \upsilon_3 + 3\upsilon_5)
+ \upsilon_3^3
\big]
\Big).
\]

Owing to the linearized dynamics and the Gaussian character of the noise operators, the steady-state fluctuations of the system are described by a Gaussian state.

\section{\label{sec3} COVARIANCE MATRIX}

Due to the linearized dynamics and the Gaussian nature of the quantum noise, the steady state of the quadrature fluctuations is a three-mode Gaussian state fully characterized by a $6 \times 6$ covariance matrix (CM) $\mathcal{V}$. Its elements are defined as
$\mathcal{V}_{\alpha\beta} = \frac{1}{2} \langle \mu_{\alpha}(t)\mu_{\beta}(t') + \mu_{\beta}(t')\mu_{\alpha}(t) \rangle$,
with $\alpha, \beta = 1,2,\dots,6$. The dynamic-state CM can be obtained by solving the corresponding Lyapunov equation~\cite{Parks1993,Vitali2007}
\begin{equation}
\partial_{t} \mathcal{V}=\mathcal{S}\mathcal{V} + \mathcal{V}\mathcal{S}^T +\mathcal{D},
\end{equation}
where $\mathcal{D} = \text{Diag}[k_x(2N_x + 1), k_x(2N_x + 1), k_a(2N_a + 1), k_a(2N_a + 1), k_d(2N_d + 1), k_d(2N_d + 1)]$ is the diffusion matrix, with its entries defined via $\mathcal{D}_{\alpha\beta} \delta(t - t') = \frac{1}{2}\langle \xi_{\alpha}(t)\xi_{\beta}(t') + \xi_{\beta}(t')\xi_{\alpha}(t) \rangle$. Here, the system is assumed to be initially prepared in the vacuum state. The covariance matrix $\mathcal{V}$ is then expressed as 
\begin{equation}
\mathcal{V}=
\begin{pmatrix}
\mathcal{Z}&\mathcal{J}\\
\mathcal{J}^{\dagger}&\mathcal{W}
\end{pmatrix},
\end{equation}
where $\mathcal{Z}$ and $\mathcal{W}$ denote the $2 \times 2$ submatrices associated with each individual mode, while the $2 \times 2$ matrix $\mathcal{J}$ describes the inter-mode cross-correlations between the two modes.

\section{\label{sec4}QUANTUM AND CLASSICAL FISHER INFORMATION for Gaussian states}
\subsection{Quantum multiparameter estimation \label{sec4.1}}

Multiparameter estimation represents a fundamental problem in quantum estimation theory, generalizing the framework of single-parameter sensing to more intricate quantum systems. In a typical quantum estimation protocol, a probe state is first prepared and subsequently evolves through a parameter-encoding quantum channel. The unknown parameters are then inferred from measurements performed on the output state using an appropriate estimator.
We consider a family of parameter-dependent quantum states $\varrho_{\boldsymbol{\tau}}$, where $\boldsymbol{\tau} = (\tau_1, \tau_2, \dots, \tau_m)$ denotes a vector composed of $m$ physical parameters to be estimated. The primary goal of quantum estimation theory is to infer the parameter vector $\boldsymbol{\tau}$ through a measurement strategy $\{\mathsf{M}_{\boldsymbol{y}}\}$ and a corresponding estimator $\boldsymbol{\hat{\tau}}(\boldsymbol{y})$. Here, the measurement process is described by a set of positive operator-valued measure (POVM) elements $\mathsf{M}_{\boldsymbol{y}}$, satisfying the completeness relation $\sum_{\boldsymbol{y}}\mathsf{M}_{\boldsymbol{y}}^{\dagger}\mathsf{M}_{\boldsymbol{y}}=\mathbb{I}$.
According to the Born rule, the probability of obtaining a measurement outcome $\boldsymbol{y}$ for a given parameter configuration $\boldsymbol{\tau}$ is expressed as
\begin{equation}
p(\boldsymbol{y}|\boldsymbol{\tau})
=
\mathrm{Tr}
\left[
\varrho_{\boldsymbol{\tau}}
\mathsf{M}_{\boldsymbol{y}}
\right].
\end{equation}

To quantify the sensitivity of the quantum state to parameter variations, we introduce the Symmetric Logarithmic Derivative (SLD) and Right Logarithmic Derivative (RLD) operators corresponding to each parameter $\tau_\alpha$. These operators are respectively defined by the following expressions \cite{EPJD2014, hmz:27, Safranek2018, Bakmou2020, Razhin2017, hmz:27}
\begin{align}
\partial_{\boldsymbol{\tau}_\alpha}\hat{\varrho}  &= \frac{\left\{\hat{\varrho},{\rm L}_\alpha^{(S)}\right\}}{2} \quad {(\rm SLD)}, \label{eq:SLD} \\
\partial_{\boldsymbol{\tau}_\alpha}\hat{\varrho} &=\hat{\varrho}  {\rm L}_\alpha^{(R)}  \quad \quad \quad {(\rm RLD)}, \label{eq:RLD}
\end{align}
 where $\partial_{\boldsymbol{\tau}_\alpha}=\frac{\partial}{\partial \boldsymbol{\tau}_\alpha}$ represents the partial derivative, where $\{\mathcal{X},\mathcal{Y}\}$ denotes the anticommutator of $\mathcal{X}$ and $\mathcal{Y}$. Based on the SLD and RLD operators, the corresponding quantum Fisher information matrices are defined as follows
\begin{align}
{\bf \mathcal{A}}_{\tau_{\alpha}\tau_{\beta}} &=  {\rm Tr} \left[ \hat{\varrho}_{\boldsymbol{\tau}}\frac{ \left\{{\rm L}^{(S)}_{\tau_{\alpha}}, {\rm L}^{(S)}_{\tau_{\beta}}\right\}}{2} \right ]\:,  \\
{\bf \mathcal{B}}_{\tau_{\alpha}\tau_{\beta}} &= {\rm Tr}\left[ \hat{\varrho}_{\boldsymbol{\tau}} {\rm L}_{\tau_{\beta}}^{(R)} {\rm L}_{\tau_{\alpha}}^{(R)\dag}\right],
\end{align}
where $\mathrm{Tr}[\Psi]$ denotes the trace operation over the finite-dimensional matrix space associated with $\Psi$. The achievable precision of parameter estimation is constrained by two distinct forms of the Cram\'er--Rao bound. In particular, by introducing the covariance matrix of the estimation outcomes,
$
\text{Cov}[\tau_{\alpha},\tau_{\beta}]
=
{\rm E}( \tau_\alpha \tau_\beta )
-
{\rm E}( \tau_\alpha )
{\rm E}( \tau_\beta ),
$
 the following inequalities hold
\begin{align}
\text{Cov}[\hat{\tau}]
&\geq \frac{{\bf \mathcal{A}}^{-1}}{\mathcal{M}},
\label{M} \\
\text{Cov}[\hat{\tau}]
&\geq \frac{{\rm Re}({\bf \mathcal{B}}^{-1})
+ \left|{\rm Im}({\bf \mathcal{B}}^{-1})\right|}{\mathcal{M}},
\label{S}
\end{align}
where $\mathcal{M}$ represents the number of measurement repetitions (or equivalently, the sample size), these inequalities simplify to lower bounds on the sum of the variances of the estimated parameters.\\
In particular, under the individual estimation strategy, the off-diagonal contributions vanish, i.e., $\mathcal{A}_{\tau_{\alpha\beta}} = \mathcal{B}_{\tau_{\alpha\beta}} = 0$ for $\alpha \neq \beta$. This implies that each parameter is independently characterized by its variance, and therefore Eqs.~(\ref{M}) and (\ref{S}) reduce to
\begin{align}
\mathrm{Var}[\tau]
&\geq \frac{{\bf \mathcal{A}}^{-1}_{\tau_{\alpha}\tau_{\alpha}}}{\mathcal{M}},
\label{M1} \\
\mathrm{Var}[\tau]
&\geq \frac{
\mathrm{Re}\!\left({\bf \mathcal{B}}^{-1}_{\tau_{\alpha}\tau_{\alpha}}\right)
+
\left|
\mathrm{Im}\!\left({\bf \mathcal{B}}^{-1}_{\tau_{\alpha}\tau_{\alpha}}\right)
\right|
}{\mathcal{M}}.
\label{S1}
\end{align}
Equations (\ref{M1}) and (\ref{S1}) are always saturable. This saturation corresponds to an optimal measurement of the parameter, where the optimal states are given by projections onto the eigenbasis of the symmetric logarithmic derivative (SLD). Furthermore, applying the trace operator to the two inequalities in (\ref{M}) and (\ref{S}) shows that they reduce to the sum of the variances of the estimated parameters.
\begin{align}
\label{BS} \sum_\alpha^{m} {\rm Var} [\tau_\alpha] &\geq { B}_{ S}/\mathcal{M} := \frac{{\bf \mathcal{A}}^{-1}}{\mathcal{M}} \:,\\
\label{RD}\sum_\alpha ^{m}{\rm Var} [\tau_\alpha] &\geq { B}_{ R}/\mathcal{M} := \frac{{\rm Re}({\bf \mathcal{B}}^{-1}) + |{\rm Im}({\bf \mathcal{B}}^{-1})| }{\mathcal{M}} \:,
\end{align}
where the inverse of the quantum Fisher information matrix has been decomposed into its real and imaginary components as $\mathcal{B}^{-1} \equiv {\rm Re}(\mathcal{B}^{-1})+ i{\rm Im}(\mathcal{B}^{-1})$, and the operator absolute value is defined by $|\Psi| \equiv \sqrt{\Psi\Psi^{\dagger}}$. Here, $\mathcal{M}$ denotes the total number of measurement repetitions. The matrices $\mathcal{A}$ and $\mathcal{B}$ correspond to the SLD \cite{hmz:11} and RLD \cite{Nichols2018, hmz:13, hmz:14} quantum Fisher information matrices, respectively.

In general, the attainability of the SLD and RLD Cram\'er--Rao bounds is subject to significant restrictions \cite{hmz:15}. Owing to the non-commutative nature of quantum measurements, achieving the ultimate precision limits for all parameters simultaneously is often impossible, since the optimization of one parameter may inevitably perturb the estimation of the others. Moreover, the optimal estimator associated with the RLD bound does not necessarily correspond to a physically implementable positive operator-valued measure (POVM). Nevertheless, even when the optimal measurements for individual parameters are incompatible, both bounds may still be saturated through appropriately designed joint measurement strategies.

Although most studies in quantum multiparameter estimation primarily focus on the SLD bound \cite{hmz:16, hmz:17, hmz:18, hmz:19, hmz:20, hmz:21, hmz:22}, it is well known that, in the single-parameter scenario, the SLD quantum Fisher information is generally smaller than its RLD counterpart. Consequently, the SLD framework provides a tighter sensitivity bound \cite{hmz:27, hmz:10}. Furthermore, while the SLD bound for single-parameter estimation is asymptotically attainable in the limit of large $\mathcal{M}$, the multiparameter setting is subject to additional constraints. Recent advances in quantum local asymptotic normality have demonstrated that the bound in Eq.~(\ref{M}) becomes asymptotically saturable if and only if the weak compatibility condition is fulfilled \cite{hmz:16},
$
\mathrm{Tr}
\left[
\varrho
\left[
{\rm L}_{\tau_\alpha},
{\rm L}_{\tau_\beta}
\right]
\right]
=0.
$
When this condition is not satisfied, the RLD bound may offer a more restrictive precision limit and thus becomes particularly relevant for accurately characterizing the estimation performance of the system.

It is therefore of fundamental importance to determine the conditions under which these inequalities can be saturated in multiparameter estimation scenarios, thereby enabling the implementation of optimal measurement strategies. In this context, it is noteworthy that several studies in quantum multiparameter estimation theory \cite{hmz:23, hmz:24, hmz:25, hmz:26}, primarily based on the SLD formalism, have established that the Quantum Cram\'er--Rao Bounds (QCRBs) given in Eqs.~(\ref{M}) and (\ref{BS}) are asymptotically attainable if and only if
\begin{equation}
{\rm Tr}\left[{\varrho \left[ {{\rm L}_{{\tau_\alpha }}^{(S)},{\rm L}_{{\tau _\beta }}^{(S)}} \right]} \right] = 0 \label{23}.
\end{equation}
It is straightforward to show that condition (\ref{23}) can be reformulated as
\begin{equation}
{\rm{Im}}\left( {{\rm Tr}\left[ { \varrho {\rm L}_{{\tau_\alpha}}^{(S)} {\rm L}_{{\tau_\beta}}^{(S)}} \right]} \right) = 0. \label{24}
\end{equation}
It is therefore pertinent to explore the link between the RLD and SLD bounds to identify which is more fundamental. This led to the development of the most informative QCRB ($B_{\rm MI}$) \cite{hmz:27, hmz:28}, defined as follows 
\begin{equation}
{B_{\rm MI}} = \max\left\{ {{B_R},{B_S}} \right\}. \label{25}
\end{equation}
Accordingly, the most informative QCRB is determined by comparing the SLD and RLD limits. We thus define the following ratio 
\begin{equation}
\chi = \frac{{{B_S}}}{{{B_R}}}.
\end{equation}
Depending on the ratio $\chi$, $B_{\rm MI}$ takes the following values
$$B_{\rm MI} = \begin{cases} B_S & \text{if } \chi < 1, \\ B_R & \text{if } \chi > 1,\end{cases}$$
naturally, if $\chi = 1$, then $B_{\rm MI} = B_R = B_S$.

In summary, the optimal precision limits in the multiparameter regime can be unified into a single inequality, expressed as
\begin{equation}
\sum\limits_\alpha^m {{\mathop{\rm var}} \left[ {{\tau_\alpha}} \right] \ge \frac{{{B_{\rm MI}}}}{\mathcal{M}}}.
\end{equation}
The SLD and RLD quantum Fisher information matrices are derived in detail in \cite{hmz:28,Bakmou2020}. Specifically, provided that $\Gamma = 2\mathcal{V} + i\Omega$ is invertible, where $\Omega=\oplus_{k=1}^n i\,\sigma_y$, with $\sigma_y$ being a Pauli matrix. The RLD matrix can be expressed as
\begin{equation}
{{\mathcal{B}}_{{\tau_\alpha}{\tau_\beta}}} = 2\mathtt{vec}{\left[ {{\partial _{{\tau_\alpha}}}\mathcal{V}} \right]^\dag }{\Sigma ^{ - 1}}\mathtt{vec}\left[ {{\partial _{{\tau_\beta}}}\mathcal{V} } \right] + 2{\partial _{{\tau_\alpha }}}{\mathbf{d}^\intercal}\hspace{0.1cm}\Gamma ^{-1}\hspace{0.1cm}{\partial _{{\tau _\beta }}}\mathbf{d}. \label{RLD}
\end{equation}
With $ {\bf{d}} = [\langle  X_x \rangle, \langle  Y_x \rangle,\langle  X_a \rangle, \langle  Y_a \rangle,\langle  X_d \rangle, \langle  Y_d \rangle]^\intercal$,  Under this condition, the Moore-Penrose pseudoinverse of $\Gamma$ reduces to its standard inverse, with $\Sigma=\Gamma^{\dag} \oplus\Gamma$. The components of the SLD quantum Fisher information matrix are given by \cite{EPJD2014, hmz:27, Bakmou2020}
\begin{equation}
{\mathcal{A}_{{\tau_\alpha }{\tau_\beta }}} = 2\mathtt{vec}{\left[ {{\partial _{{\tau_\alpha}}}\mathcal{V} } \right]^\dag }{{\mathcal P}^{+}}\mathtt{vec}\left[ {{\partial _{{\tau_\beta}}}\mathcal{V} } \right] + {\partial _{{\tau_\alpha}}}{{\bf{d}}^\intercal} \hspace{0.1cm} \mathcal{V}^{-1} \hspace{0.1cm}{\partial _{{\tau_\beta}}}{\bf{d}},\label{37}
\end{equation}
with $\mathcal{P} = 4{\mathcal{V}^\dagger \otimes \mathcal{V} + \Omega \otimes \Omega}$. Provided that $\mathcal{P}$ is invertible, the SLD quantum Fisher information matrix can be expressed as
\begin{equation}
{\mathcal{A}_{{\tau_\alpha }{\tau_\beta }}} = 2\mathtt{vec}{\left[ {{\partial _{{\tau_\alpha}}}\mathcal{V} } \right]^\dag }{{\mathcal P}^{-1}}\mathtt{vec}\left[ {{\partial _{{\tau_\beta}}}\mathcal{V} } \right] + {\partial _{{\tau_\alpha}}}{{\bf{d}}^\intercal} \hspace{0.1cm} \mathcal{V}^{-1} \hspace{0.1cm}{\partial _{{\tau_\beta}}}{\bf{d}}.\label{SLD}
\end{equation}
Thus, for the estimation of the parameters: the exciton-cavity coupling $g$ or dissipation rate of exciton mode $k_x$, the RLD quantum Fisher information matrix $\mathcal{B}$, can be calculated from Eq. (\ref{RLD}), as
\begin{equation}
\mathcal{B} =\begin{pmatrix}
2\mathtt{vec}{\left[ {{\partial _{g}}\mathcal{V}} \right]^\dag }{\Sigma ^{ - 1}}\mathtt{vec}\left[ {\partial_{g}}\mathcal{V} \right] + 2{\partial _{g}}{\mathbf{d}^\intercal}\hspace{0.1cm}\Gamma ^{-1}\hspace{0.1cm}{\partial _{g}}\mathbf{d} & 2\mathtt{vec}{\left[ {{\partial _{g}}\mathcal{V}} \right]^\dag }{\Sigma^{ - 1}}\mathtt{vec}\left[ {{\partial_{{ k_{x}}}}\mathcal{V} } \right] + 2{\partial _{g}}{\mathbf{d}^\intercal}\hspace{0.1cm}\Gamma ^{-1}\hspace{0.1cm}{\partial _{{ k_{x}}}}\mathbf{d} \\
2\mathtt{vec}{\left[ {{\partial _{{ k_{x}}}}\mathcal{V}} \right]^\dag }{\Sigma ^{ - 1}}\mathtt{vec}\left[ {{\partial_{g}}\mathcal{V} } \right] + 2{\partial _{{ k_{x}}}}{\mathbf{d}^\intercal}\hspace{0.1cm}\Gamma ^{-1}\hspace{0.1cm}{\partial _{g}}\mathbf{d} & 2\mathtt{vec}{\left[ {{\partial _{k_{x}}}\mathcal{V}} \right]^\dag }{\Sigma ^{ - 1}}\mathtt{vec}\left[ {{\partial_{k_{x}}}\mathcal{V} } \right] + 2{\partial _{k_x}}{\mathbf{d}^\intercal}\hspace{0.1cm}\Gamma ^{-1}\hspace{0.1cm}{\partial _{k_x}}\mathbf{d}
\end{pmatrix}, \label{RLDM}
\end{equation}
the SLD quantum Fisher information matrix $\mathcal{A}$, can be calculated from Eq. (\ref{SLD}), as
\begin{equation}
\mathcal{A} =\begin{pmatrix}
2\mathtt{vec}{\left[ {{\partial _{g}}\mathcal{V}} \right]^\dag }{\mathcal{P}^{ - 1}}\mathtt{vec}\left[ {{\partial_{g}}\mathcal{V} } \right] + {\partial _{g}}{\mathbf{d}^\intercal}\hspace{0.1cm}\mathcal{V}^{-1}\hspace{0.1cm}{\partial _{g}}\mathbf{d} & 2\mathtt{vec}{\left[ {{\partial _{g}}\mathcal{V}} \right]^\dag }{\mathcal{P}^{ - 1}}\mathtt{vec}\left[ {{\partial_{{k_x}}}\mathcal{V} } \right] + {\partial _{g}}{\mathbf{d}^\intercal}\hspace{0.1cm}\mathcal{V}^{-1}\hspace{0.1cm}{\partial _{k_x}}\mathbf{d} \\
2\mathtt{vec}{\left[ {{\partial _{k_x}}\mathcal{V}} \right]^\dag }{\mathcal{P} ^{ - 1}}\mathtt{vec}\left[ {{\partial_{g}}\mathcal{V} } \right] + {\partial _{k_x}}{\mathbf{d}^\intercal}\hspace{0.1cm}\mathcal{V}^{-1}\hspace{0.1cm}{\partial _{g}}\mathbf{d} & 2\mathtt{vec}{\left[ {{\partial _{k_x}}\mathcal{V}} \right]^\dag }{\mathcal{P}^{ - 1}}\mathtt{vec}\left[ {{\partial_{k_{x}}}\mathcal{V} } \right] + {\partial _{k_x}}{\mathbf{d}^\intercal}\hspace{0.1cm}\mathcal{V}^{-1}\hspace{0.1cm}{\partial _{k_x}}\mathbf{d}
\end{pmatrix}. \label{SLDM}
\end{equation}
Then, $B_S$ can be calculated from Eqs. (\ref{BS}) and (\ref{RLDM}), and $B_R$ from Eqs. (\ref{RD}) and (\ref{SLDM}).

\subsection{Classical Fisher information} \label{IIIB}

This section presents the framework of the Classical Fisher Information (CFI). Within this formalism, a quantum measurement is described by a Positive Operator-Valued Measure (POVM) $\{\hat{\Pi}_{y}\}$, where each measurement operator is positive semi-definite, i.e., $\hat{\Pi}_{y} \geq 0$, and the set fulfills the completeness relation $\sum_{y}\hat{\Pi}^{\dagger}_y\hat{\Pi}_{y} = \mathbb{I}$ \cite{hmz:30}. The collection of all admissible measurement strategies is denoted by $\widetilde{\Omega}$. According to the Born rule, the measurement outcomes are governed by the conditional probability distribution
\begin{equation}
\{P(y|\tau)=\text{Tr}[{\varrho}_{\tau}\hat{\Pi} _{y}]\}, 
\end{equation}%
where $P(y|\tau)$ represents the conditional probability associated with obtaining the measurement outcome $y$ for a given parameter value $\tau$. Based on this probability distribution, the parameter $\tau$ can be estimated, and the corresponding Classical Fisher Information (CFI) is defined as \cite{hmz:11,hmz:31,hmz:32}
\begin{equation}
\label{Eq24}
\mathcal{F}_{\tau}=\int \frac{1}{P(y|\tau)}\left[ \frac{\partial P(y|\tau)}{\partial \tau}\right] ^{2}dy.
\end{equation}
Since the QFI is defined as $\mathcal{A}_{\tau} := \max_{\tilde{\Omega}} \{\mathcal{F}_{\tau}\}$, it inherently accounts for the optimal measurement strategy. Thus, we obtain
\begin{equation}
\text{Var}\left(\hat{\tau}\right) \geq \frac{1}{\mathcal{M}\mathcal{F}_{\tau}}\geq \frac{1}{\mathcal{M}\mathcal{A}_{\tau}}.
\end{equation}
The lower bound of the classical CRB is achievable through optimal estimation strategies; for instance, the maximum likelihood estimator is known to be asymptotically efficient. It should be noted that all measurements considered in this work are assumed to be ideal, corresponding to a measurement efficiency of unity. The calculation of the CFI is performed using the second-moment matrix $\mathcal{V}$ and the first-moment vector $\mathbf{d}$.

\noindent{\bf Heterodyne detection.}  Heterodyne detection is a fundamental Gaussian measurement technique in which the signal field is mixed with a reference (local oscillator) field of different frequency. The associated measurement operators are described by the overcomplete set of coherent-state projectors $\{|\alpha\rangle\langle\alpha|/\pi\}$, which form a valid positive operator-valued measure (POVM) \cite{hmz:10,Meystre2007,Scully1999,hmz:35}.  In practice, heterodyne detection is implemented by combining the mode of interest with an ancillary vacuum mode at a balanced $50{:}50$ beam splitter, followed by simultaneous measurements of the conjugate quadratures $X$ and $Y$ of the output modes. For the cavity–exciton system, a detailed realization of optimal heterodyne detection on the relevant subsystems is provided in \cite{hmz:36}.
\begin{equation}
\label{Eq33}
\mathcal{F}_{\tau,\text{Het}}=\frac{1}{2}\text{Tr}\bigg(\psi^{-1}\frac{d\psi}{d\tau}\psi^{-1}\frac{d\psi}{d\tau}\bigg) + \frac{d\mathbf{d}^{\intercal}}{d\tau}\psi^{-1}\frac{d\mathbf{d}}{d\tau},
\end{equation}
where $\psi = \mathcal{V} + \mathbb{I}_{6\times 6}$, and the subscript “Het”  marks the heterodyne detection. Here, $\mathbf{d}$ and $\mathcal{V}$ are the first and second moments of the bipartite system, while $\mathbb{I}_{6\times 6}$ represents the supplemental noise inherent in the simultaneous measurement of conjugate quadratures $X$ and $Y$.

\noindent{\bf Homodyne detection.} In homodyne detection, the signal field is combined with a strong local oscillator at the same frequency, enabling the measurement of a single field quadrature. The corresponding measurement is described by the POVM ${|Q\rangle\langle Q|}$ or ${|P\rangle\langle P|}$, where $|Q\rangle$ and $|P\rangle$ denote the eigenstates of the quadrature operators $Q$ and $P$, respectively \cite{hmz:10,Meystre2007,Scully1999,hmz:35}. Here, $Q \in \{X_x, X_a\}$ and $P \in \{Y_x, Y_a\}$; the cavity quadratures are directly accessible, whereas the exciton quadratures are inferred indirectly via auxiliary field measurements. Experimentally, this is typically realized by coupling a exciton to a weakly driven microwave cavity. Within the homodyne detection scheme, the classical Fisher information (CFI) associated with the parameter $\tau$ is given by \cite{Monras2013}.
\begin{equation}
\label{Eq31}
\mathcal{F}_{\tau,\text{Ho}}^{l}=\frac{1}{2\mathcal{V}_{ll}^{2}}\left[ 2\mathcal{V}_{ll}(\partial_{\tau} \mathbf{d}_{l})^{2}+\left( \partial_{\tau}\mathcal{V}_{ll}\right) ^{2}\right],
\end{equation}%
where, the superscript '$l$' labels the measured quadrature, while the subscript 'Ho' refers to homodyne detection. Specifically, the CFI for parameters such as the cavity-exciton coupling $g$ or dissipation rate of exciton mode $k_x$, utilizing homodyne measurements of $\mathbf{d}_{1} = \langle X_x \rangle$ and $\mathbf{d}_{2} = \langle Y_x \rangle$, are expressed as
\begin{equation}
\label{Eq31}
\mathcal{F}_{g,\text{Ho}}^{X_x}=\frac{1}{2\mathcal{V}_{11}^{2}}\left[ 2\mathcal{V}_{11}(\partial_{g}\langle X_x \rangle)^{2}+\left( \partial_{g}\mathcal{V}_{11}\right) ^{2}\right],\qquad \mathcal{F}_{k_x,\text{Ho}}^{ X_x}=\frac{1}{2\mathcal{V}_{11}^{2}}\left[ 2\mathcal{V}_{11}(\partial_{k_x}\langle X_x \rangle)^{2}+\left( \partial_{k_x}\mathcal{V}_{11}\right) ^{2}\right],
\end{equation}
and 
\begin{equation}
\label{Eq31}
\mathcal{F}_{g,\text{Ho}}^{ Y_{x}}=\frac{1}{2\mathcal{V}_{22}^{2}}\left[ 2\mathcal{V}_{22}(\partial_{g}\langle Y_{x} \rangle)^{2}+\left( \partial_{g}\mathcal{V}_{22}\right) ^{2}\right],\qquad \mathcal{F}_{k_x,\text{Ho}}^{ Y_{x}}=\frac{1}{2\mathcal{V}_{22}^{2}}\left[ 2\mathcal{V}_{22}(\partial_{k_x}\langle Y_{x} \rangle)^{2}+\left( \partial_{k_x}\mathcal{V}_{22}\right) ^{2}\right].
\end{equation}

\section{Results and discussions \label{sec5}}

We now investigate the sensitivity of the estimation precision for the coupling strength $g$ and the cavity decay rate $k_a$ with respect to various system parameters in both steady-state and dynamical regimes. These parameters include the environmental temperature $\rm T$, driving power $\mathrm{P}$, as well as the dissipation rates and detunings associated with the cavity and exciton modes, namely $(k_a,k_x)$ and $(\Delta_a,\Delta_x)$, respectively. Furthermore, we examine the influence of the coupling strength $G_0$ on the estimation performance. In addition, a comparison is carried out between the SLD-based quantum bound and the classical Fisher information (CFI) obtained through heterodyne and homodyne detection schemes. Unless otherwise stated, the numerical simulations are performed using experimentally feasible parameters reported in the table \ref{table} 
\begin{table}[h!]
\centering
\caption{Parameters used in numerical simulations \cite{Santos2023,Kuznetsov2023,Zuo2024} .}
\label{table}
\begin{tabular}{l c c c}
\hline\hline
\textbf{Parameter} & \textbf{Symbol} & \textbf{Value} & \textbf{Unit} \\
\hline
Phonon mode frequency            & $\Omega_d/2\pi$     & $20\times10^{9}$ & Hz \\
Phonon decay rate                & $k_d/2\pi$          & $1\times 10^{6}$         & Hz \\
Cavity decay rate                & $k_a/2\pi$          & $1\times10^{9}$         & Hz \\
Exciton decay rate               & $k_x/2\pi$          & $1\times10^{8}$         & Hz \\
Photon--phonon coupling          & $G_0/2\pi$          & $1\times10^{7}$         & Hz \\
Drive amplitude                  & $\varepsilon/2\pi$  & $6\times10^{12}$ & Hz \\
Driving power                    & $P$            & $26\times 10^{-3}$                       & W \\
Drive field frequency            & $\Omega_0/2\pi$     & $345\times10^{12}$ & Hz \\
Temperature                      & $\rm T$            & $1$                        & K \\
Exciton--photon coupling         & $g/2\pi$            & $0.9\times10^{9}$ & Hz \\
\hline\hline
\end{tabular}
\end{table}
\subsection{Steady-state}

In this section, we investigate the steady-state multiparameter estimation performance of the hybrid exciton--optomechanical system. Particular attention is devoted to the influence of experimentally relevant parameters on the most informative quantum Cram\'er--Rao bound and the achievable estimation precision.

\begin{figure}[!htb]
\includegraphics[width=0.4\linewidth]{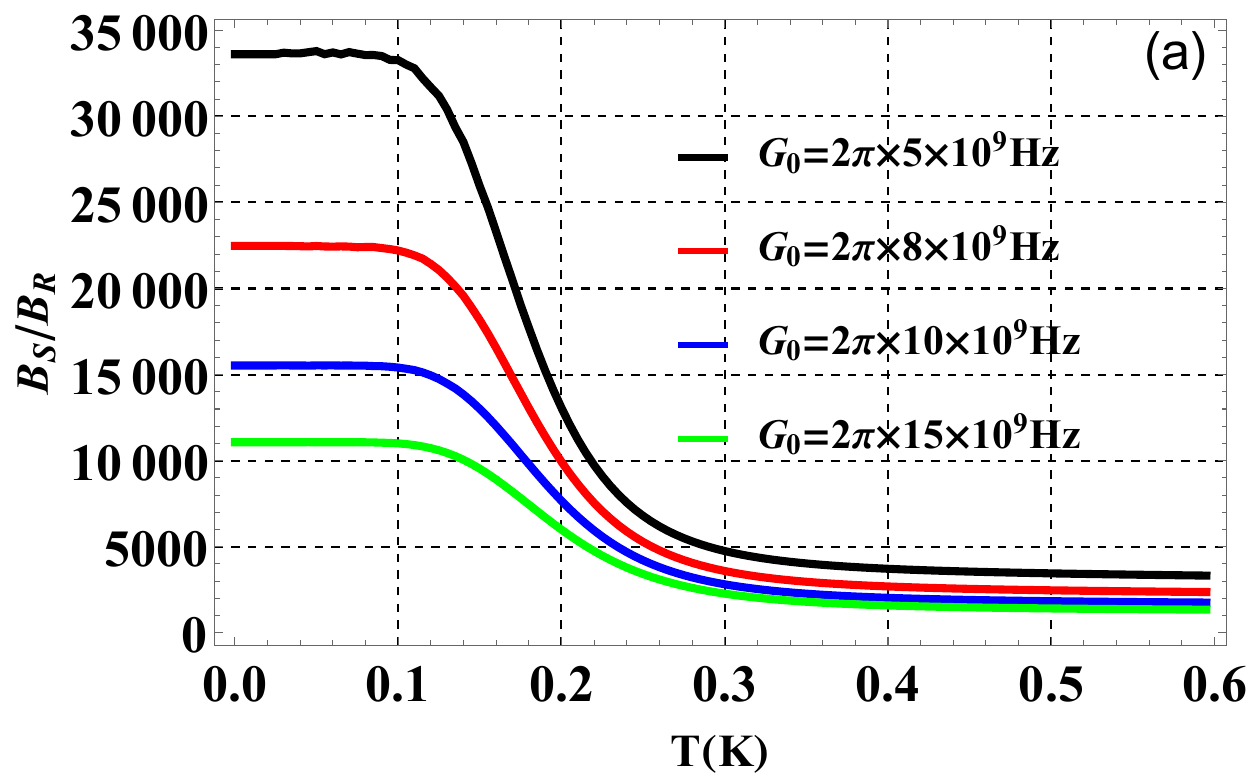}
\includegraphics[width=0.4\linewidth]{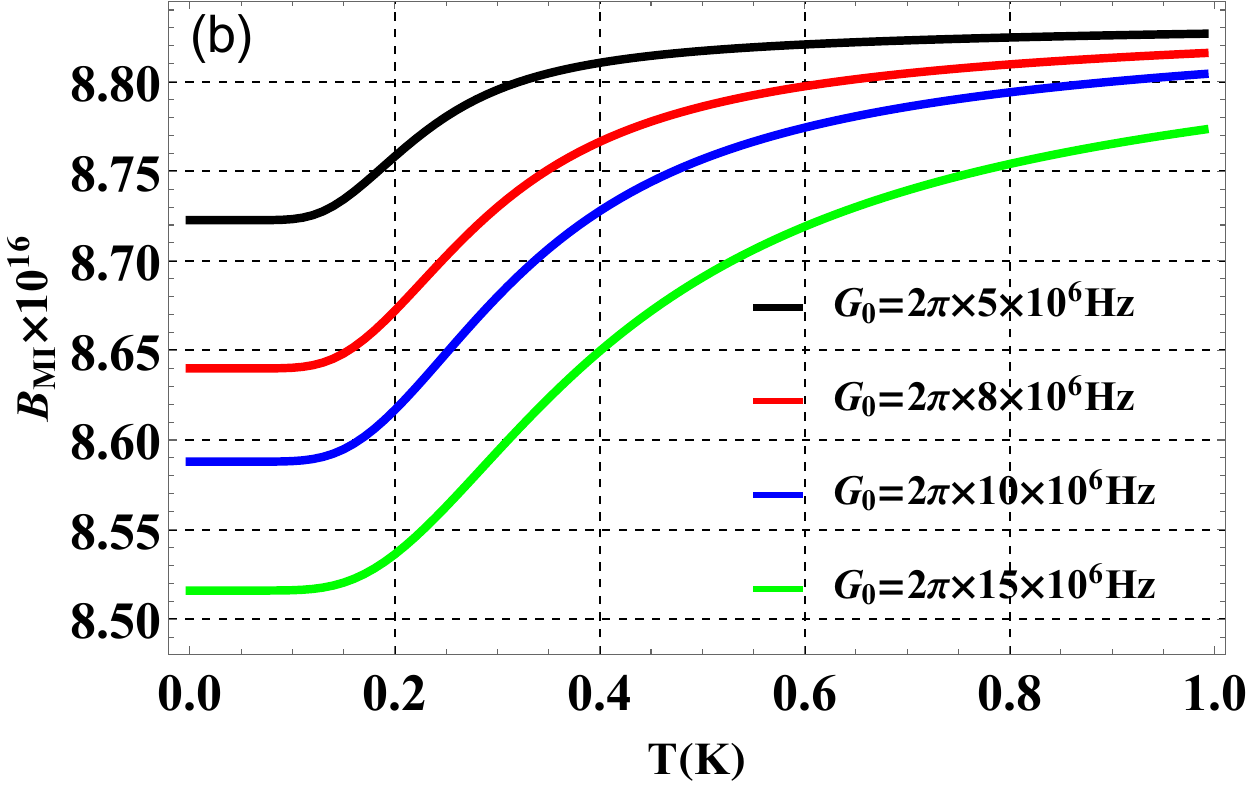}
\caption{(a) Plot of the $B_S/B_R$ as function of the temperature $\rm T$ for different values of the coupling $G_0$. (b) Plot of the $B_{\rm MI}$ as function of the temperature $\rm T$ for different values of the coupling $G_0$. The other parameters are provided in the table \ref{table}.}
\label{BSR}
\end{figure}

Fig.\ref{BSR}(a) illustrates the behavior of the ratio $\chi=B_S/B_R$ as a function of the temperature $\rm T$ for different values of the optomechanical coupling strength $G_0$. One observes that the ratio strongly decreases as the temperature increases, indicating a gradual reduction of the gap between the SLD and RLD bounds. In the low-temperature regime, the ratio attains large values, showing that the RLD bound provides a significantly tighter precision limit than the SLD formalism. Physically, this behavior originates from the strong quantum coherence preserved at low temperatures, where thermal fluctuations remain weak and the hybrid exciton-optomechanical correlations are maximized. As the temperature increases, thermal phonons progressively destroy the quantum correlations established between the excitonic, optical, and mechanical modes. Consequently, the incompatibility effects between the estimated parameters become weaker, reducing the distinction between the SLD and RLD bounds. Furthermore, increasing $G_0$ suppresses the ratio $B_S/B_R$, indicating that stronger optomechanical interactions reduce the relative advantage of the RLD estimation strategy. This behavior is attributed to the redistribution of quantum fluctuations induced by the enhanced radiation-pressure interaction.

Fig. \ref{BSR}(b) presents the variation of the most informative bound ($B_{\rm MI}$) as a function of the environmental temperature for different values of the optomechanical coupling strength $G_0$. It is observed that the $B_{\rm MI}$ increases monotonically with temperature, demonstrating a progressive degradation of the estimation precision. This behavior is physically expected since thermal noise introduces incoherent fluctuations into the hybrid system and suppresses the quantum coherence responsible for enhanced parameter sensitivity. In the low-temperature regime, the phononic thermal occupation remains weak, allowing the hybrid system to preserve strong exciton-photon-mechanical correlations. Consequently, the quantum state becomes highly distinguishable under infinitesimal variations of the estimated parameters, leading to improved precision. Moreover, increasing the optomechanical coupling $G_0$ reduces the $B_{\rm MI}$ over the whole temperature region. This enhancement originates from the stronger interaction between the cavity field and the mechanical mode, which amplifies the sensitivity of the steady-state Gaussian fluctuations to parameter changes. These results indicate that strong optomechanical coupling combined with low-temperature operation provides the optimal regime for high-precision multiparameter quantum estimation.

\begin{figure}[!htb]
\includegraphics[width=0.5\linewidth]{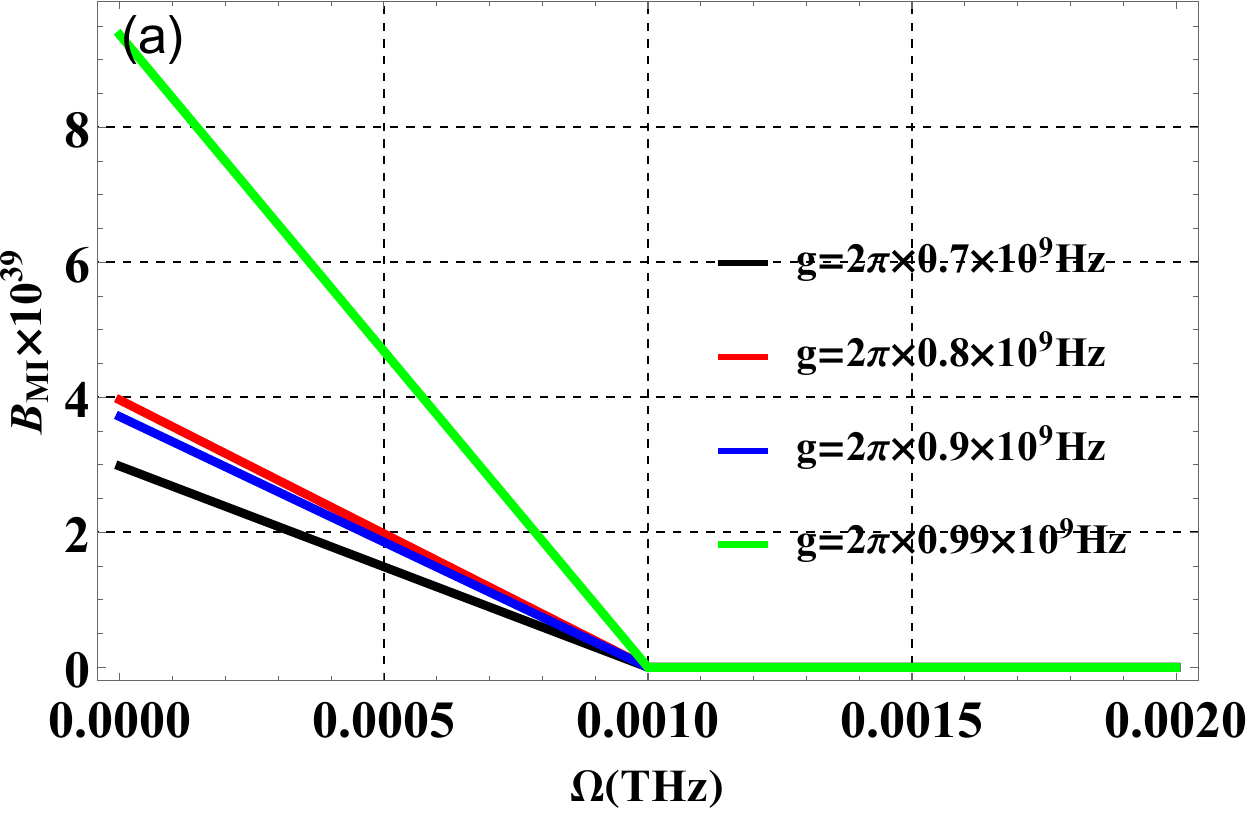}
\caption{Plot of the most informative bound ($B_{\rm MI}$) as function of the driving amplitude $\varepsilon$ for different values of the coupling $\rm g$. The other parameters are provided in the table \ref{table}.}
\label{Omega}
\end{figure}

Figure \ref{Omega} shows the dependence of the $B_{\rm MI}$ on the driving amplitude $\varepsilon$ for different values of the exciton-photon coupling strength $\rm g$. One clearly observes that the $B_{\rm MI}$ rapidly decreases as the driving amplitude increases and eventually approaches a nearly vanishing value. Physically, increasing the driving strength enhances the intracavity photon population, which effectively amplifies the hybrid interactions within the system. As a result, the encoded information associated with the estimated parameters becomes more efficiently transferred into the quantum fluctuations of the output state. Consequently, the distinguishability between neighboring quantum states is significantly enhanced, leading to a substantial improvement of the estimation precision. In addition, larger values of the coupling strength $\rm g$ further accelerate the reduction of the $B_{\rm MI}$. This behavior arises from the stronger coherent exchange of excitations between the excitonic and photonic modes, which increases the susceptibility of the system to small parameter variations. The obtained results demonstrate that the strong-driving regime constitutes an efficient operating region for achieving ultrasensitive multiparameter quantum estimation.

\begin{figure}[!htb]
\includegraphics[width=0.5\linewidth]{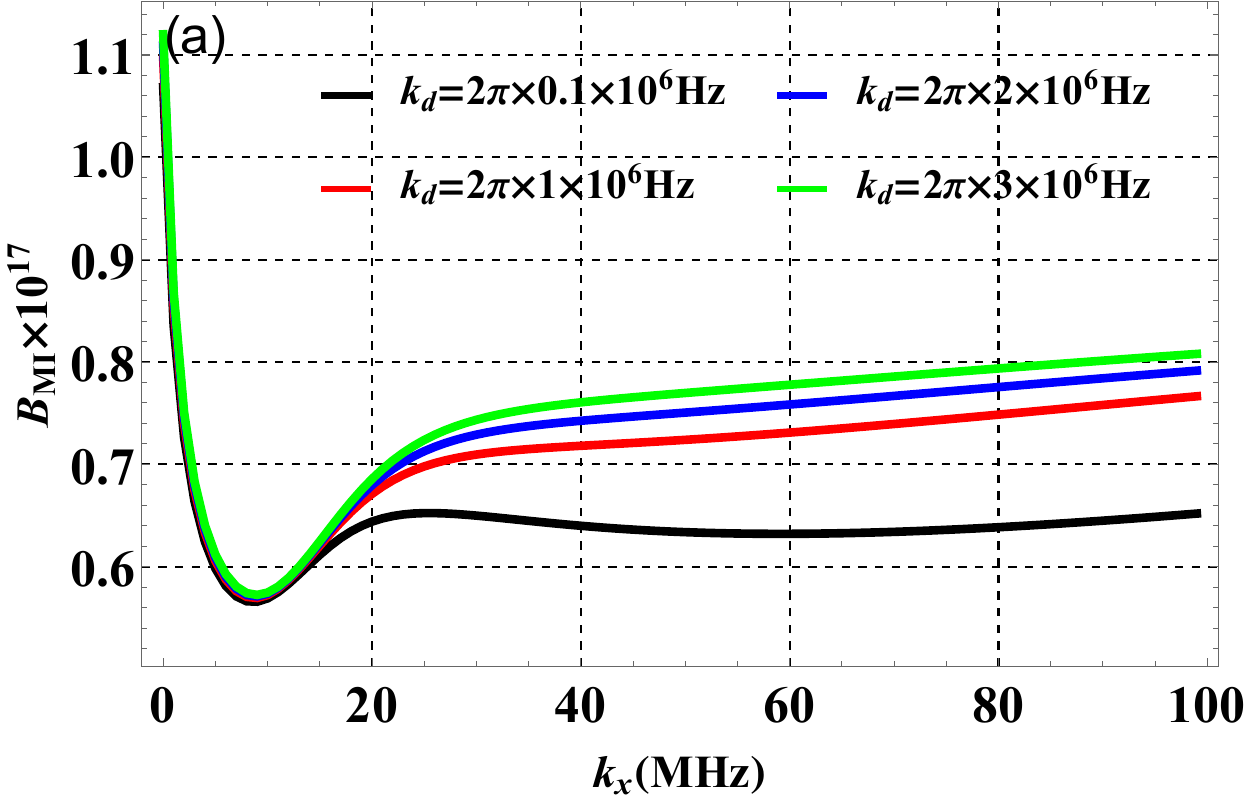}
\caption{Plot of the most informative bound ($B_{\rm MI}$) as function of the exciton decay rate $k_x$ for different values of the mechanical damping rate $k_d$. The other parameters are provided in the table \ref{table}.}
\label{kappax}
\end{figure}

Fig. \ref{kappax} displays the variation of the $B_{\rm MI}$ as a function of the exciton decay rate $k_x$ for different values of the mechanical damping rate $k_d$. One observes that the $B_{\rm MI}$ initially decreases, reaches a minimum value, and then slightly increases before approaching a nearly stationary regime. The existence of this optimal region indicates that moderate excitonic dissipation can improve the estimation precision by stabilizing the hybrid dynamics and suppressing excess fluctuation noise. Physically, weak dissipation allows the system to preserve coherent exciton-photon interactions while avoiding instability-induced fluctuation amplification. However, for larger values of $k_x$, decoherence processes become dominant and progressively suppress the useful quantum correlations responsible for high estimation sensitivity. Consequently, the distinguishability between nearby quantum states decreases, leading to the degradation of the metrological performance. Furthermore, increasing the mechanical damping $k_d$ slightly increases the $B_{\rm MI}$, showing that phonon losses negatively affect the estimation precision. This behavior confirms that mechanical dissipation acts as an additional decoherence channel that weakens the coherent transfer of information inside the hybrid system.

\subsection{Dynamical state}

We now turn to the dynamical regime and analyze the time evolution of the estimation precision. This allows us to reveal the interplay between coherent hybrid interactions and dissipative processes governing the temporal behavior of the quantum Fisher information and the corresponding precision bounds.

\begin{figure}[!htb]
\includegraphics[width=0.4\linewidth]{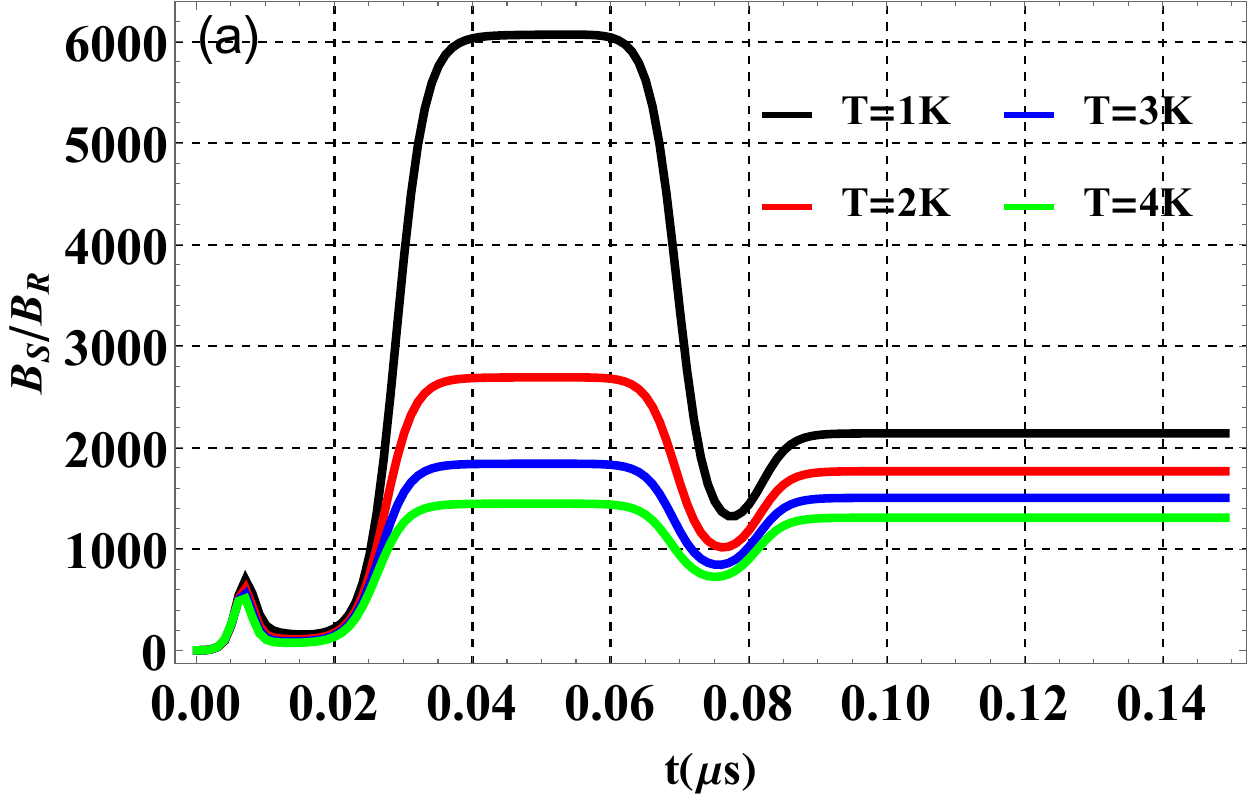}
\includegraphics[width=0.4\linewidth]{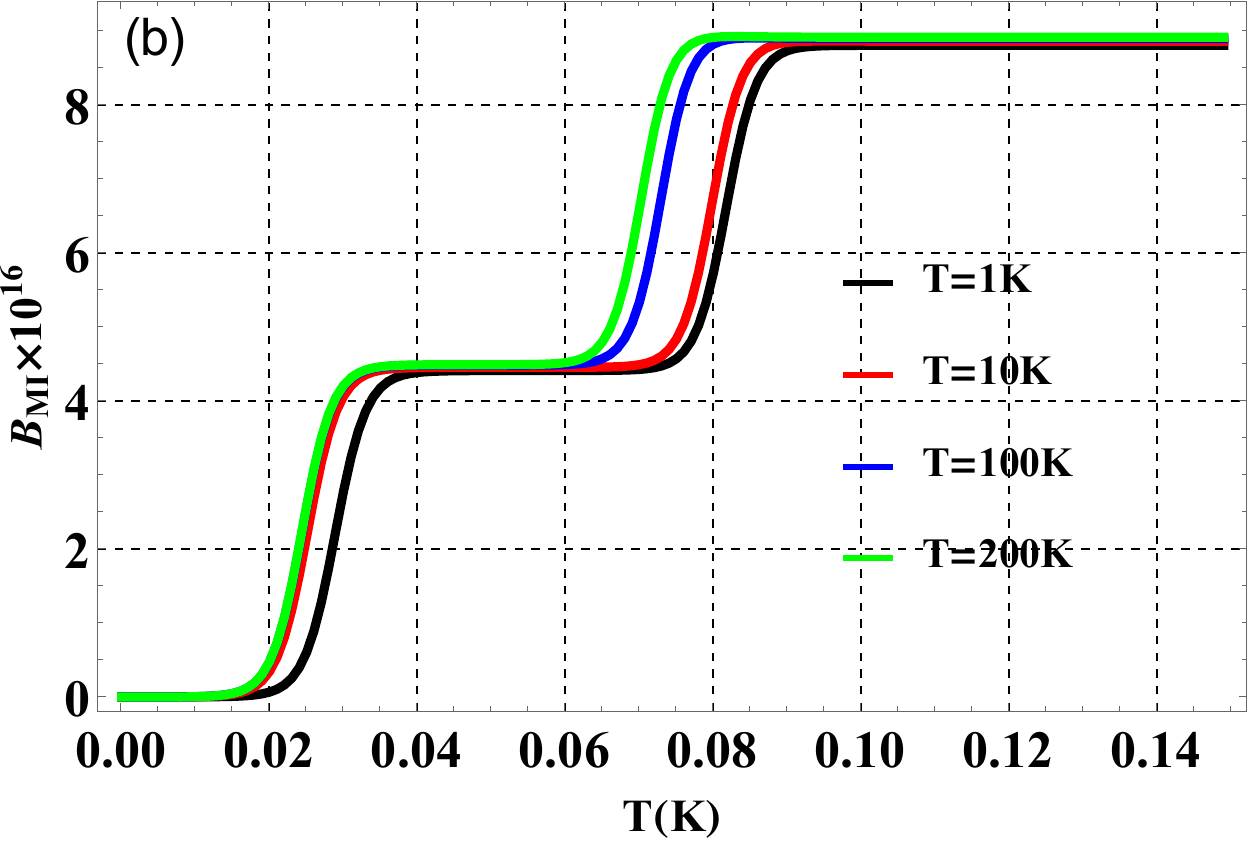}
\caption{Plot of the dynamical evolution of the : (a) the ratio $B_S/B_R$ and (b) the most informative bound ($B_{\rm MI}$) for different values of the temperature $\rm T$.  The other parameters are provided in the table \ref{table}.}
\label{BSRd}
\end{figure}

We plot in Fig. \ref{BSRd} the dynamical evolution of the ratio $\chi=B_S/B_R$ and the most informative bound ($B_{\rm MI}$) for different bath temperatures $\rm T$. As depicted in Fig.~\ref{BSRd}(a), the ratio $B_S/B_R$ exhibits a pronounced transient behavior characterized by an initial increase followed by damped oscillations before reaching a stationary state. This behavior can be originates from the competition between coherent exciton--photon--phonon interactions and dissipative processes. During the early evolution stage, quantum correlations are rapidly established among the three subsystems as discussed in \cite{Zuo2024}, leading to a strong distinction between the SLD and RLD estimation strategies. As time progresses, environmental dissipation progressively suppresses these correlations, causing the ratio $B_S/B_R$ to converge toward a steady-state value. When $\chi>1$, corresponding to $B_S>B_R$, the SLD bound becomes larger than the RLD bound, and therefore the ultimate estimation precision is determined by $B_S$. This means that the SLD formalism provides the most restrictive and informative precision limit. In contrast, when $\chi<1$, corresponding to $B_S<B_R$, the RLD bound becomes dominant and the $B_{\rm MI}$ is given by $B_R$. This means that the RLD approach more appropriate for characterizing the achievable precision. However, one can see that the distinction between the SLD and RLD bounds, i.e., $B_S>B_R$ for a wide range of time ($t>0.02\mu$s), as depicted in Fig. \ref{BSRd}(a). Besides, increasing the temperature significantly reduces the magnitude of $B_S/B_R$. This behavior reflects the detrimental role of thermal phonons, which introduce incoherent fluctuations into the mechanical resonator and weaken the quantum coherence responsible for parameter sensitivity. Consequently, the difference between the SLD and RLD bounds becomes less pronounced at higher temperatures.

Fig~\ref{BSRd}(b) explores the corresponding evolution of the $B_{\rm MI}$. One can see a step-like increase followed by saturation at long times. Since the $B_{\rm MI}$ represents the ultimate multiparameter quantum Cram\'er--Rao bound, its increase directly indicates a degradation of the attainable estimation precision. This can be explain by the thermal fluctuations reduce the distinguishability between neighboring quantum states associated with different parameter values. As a result, the encoded information about the exciton--photon coupling $\rm g$ and excitonic dissipation $k_x$ becomes progressively less accessible. Moreover, higher temperatures lead to larger stationary $B_{\rm MI}$ values, demonstrating that thermal noise constitutes the dominant mechanism limiting the metrological performance of the hybrid exciton--optomechanical system. Therefore, the optimal estimation regime is achieved at low temperatures, i.e., where quantum coherence and hybrid correlations (entanglement) remain maximally preserved.

\begin{figure}[!htb]
\includegraphics[width=0.4\linewidth]{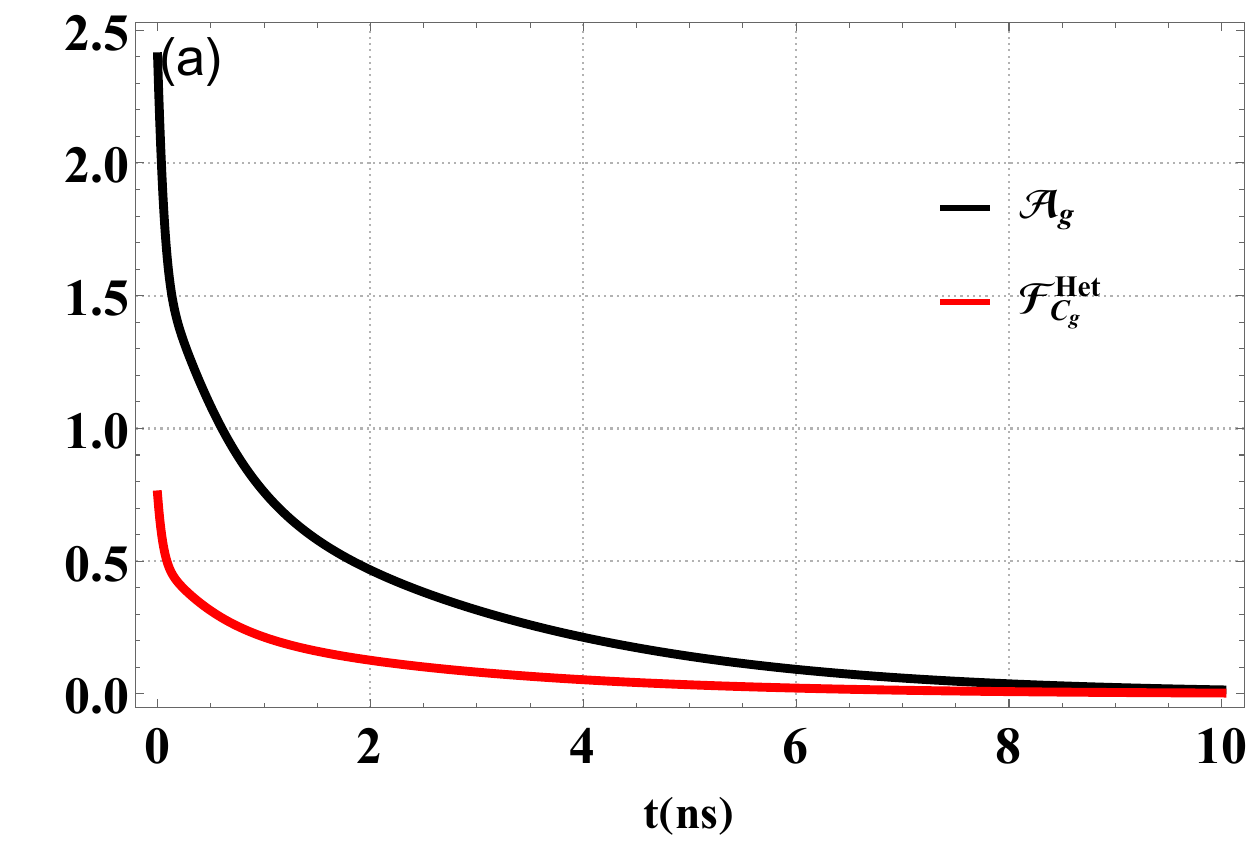}
\includegraphics[width=0.4\linewidth]{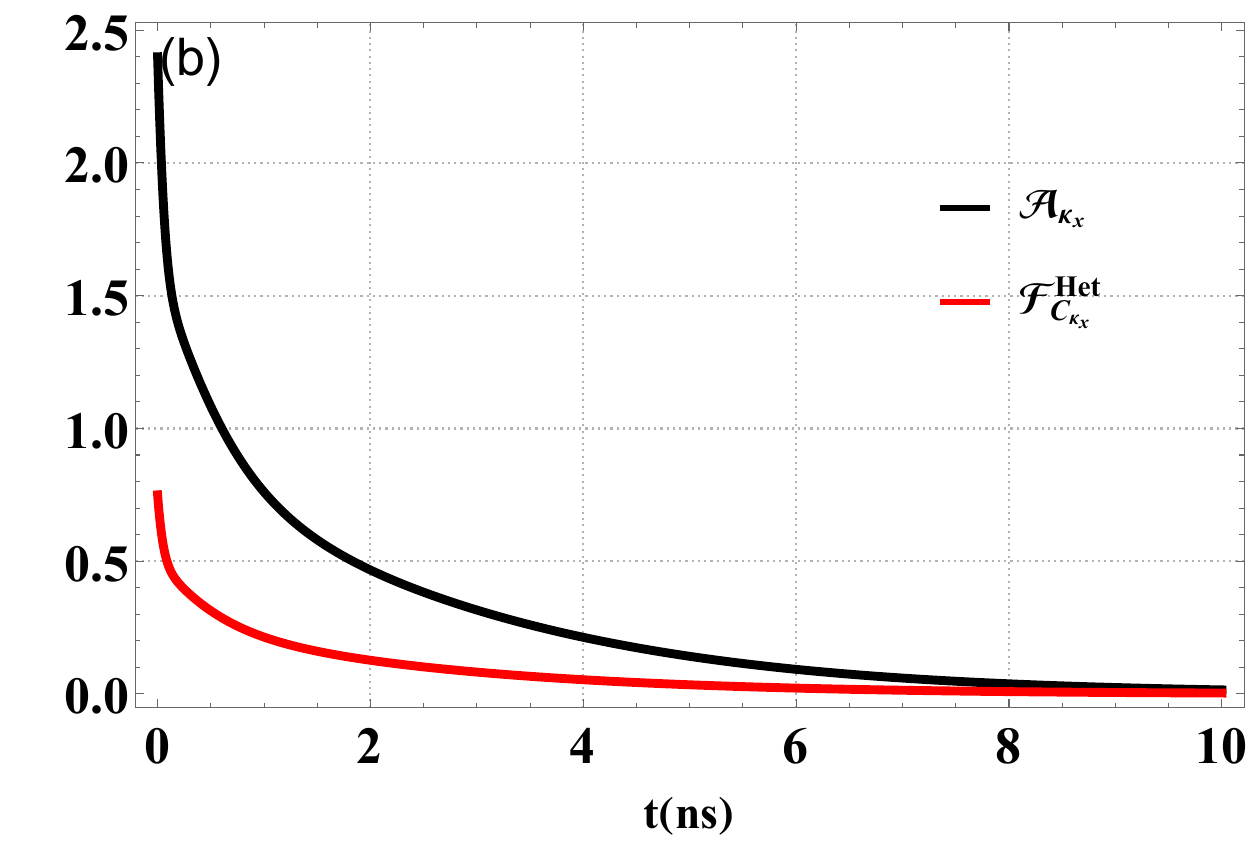}
\caption{Plot of the dynamical evolution of  (a) the quantum Fisher information $\mathcal{F}_g$ and the classical Fisher information CFI in Heterodyne detection  $\mathcal{F}^{\rm Het}_{C_g}$ and (b) the quantum Fisher information $\mathcal{F}_{k_x}$ and the classical Fisher information CFI in Heterodyne detection  $\mathcal{F}^{\rm Het}_{C_{k_x}}$. The other parameters are provided in the table \ref{table}.}
\label{QCt}
\end{figure}

Figure \ref{QCt} illustrates the dynamical evolution of the quantum Fisher information associated with the estimation of the parameters $\rm g$ and $k_x$, together with the corresponding heterodyne classical Fisher information. One observes that both the QFI and CFI rapidly decrease with time before approaching a stationary regime at longer evolution times. Initially, the quantum state contains a large amount of accessible information about the estimated parameters due to the strong coherent correlations established within the hybrid system. As the system evolves, dissipative processes progressively destroy these quantum correlations and reduce the sensitivity of the Gaussian state to infinitesimal parameter variations. Consequently, the estimation precision deteriorates with increasing interaction time. Moreover, the heterodyne CFI remains lower than the corresponding QFI during the whole evolution, indicating that heterodyne detection does not fully saturate the ultimate quantum limit. Nevertheless, the relatively close behavior between the two quantities demonstrates that heterodyne measurements still provide an efficient experimentally accessible strategy for extracting parameter information from the output field.

\begin{figure}[!htb]
\includegraphics[width=0.4\linewidth]{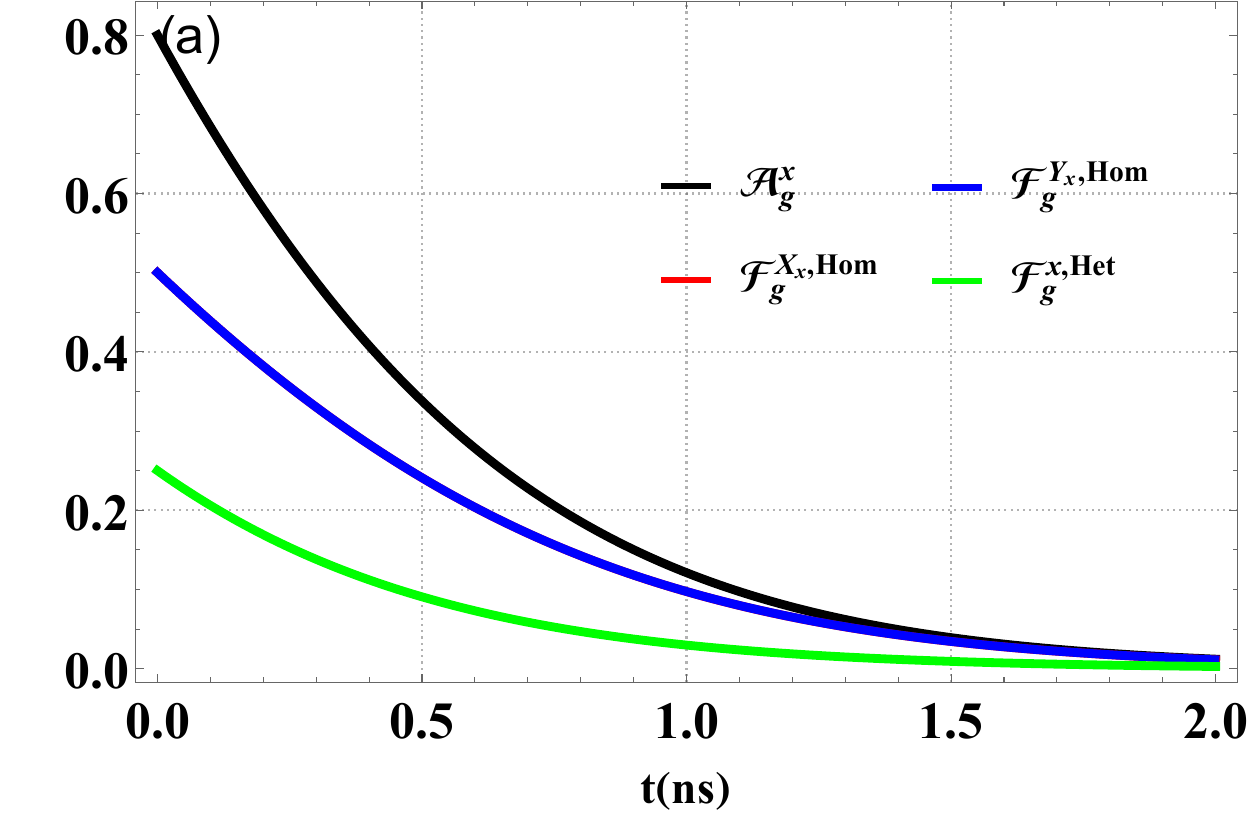}
\includegraphics[width=0.4\linewidth]{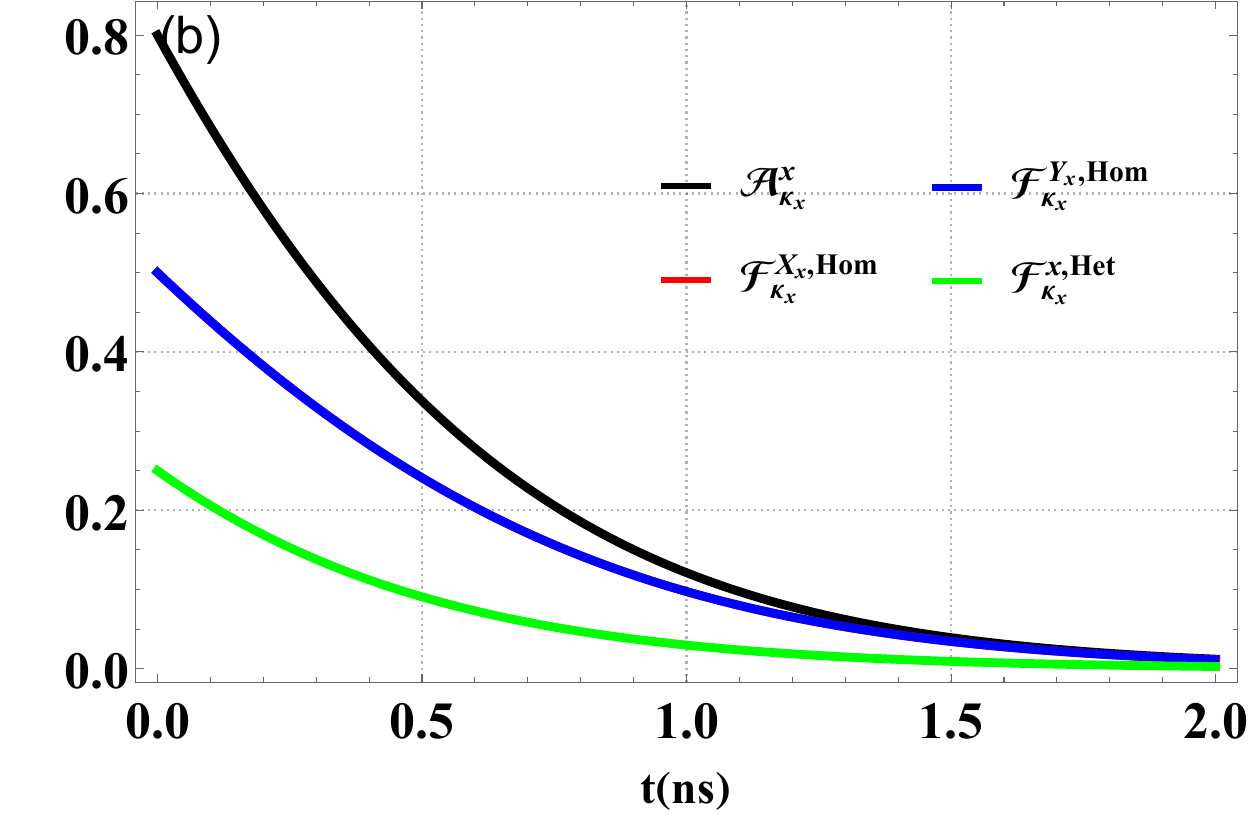}
\caption{Time evolution of (a) the SLD QFI for the exciton mode ($x$) $\mathcal{F}^{ x,{\rm SLD}}_{g}$ and the QFI in homodyne detection $\mathcal{F}^{Y_{ x},{\rm Hom}}_{g}$ and $\mathcal{F}^{X_{ x},{\rm Hom}}_{g}$ and $\mathcal{F}^{x,{\rm Hom}}_{g}$, (b) the SLD QFI for the exciton mode ($x$) $\mathcal{F}^{ x,{\rm SLD}}_{k_x}$ and the QFI in homodyne detection $\mathcal{F}^{Y_{ x},{\rm Hom}}_{k_x}$ and $\mathcal{F}^{X_{ x},{\rm Hom}}_{g}$ and $\mathcal{F}^{x,{\rm Hom}}_{g}$, for quadrature operators $X_{x}$ and $Y_{ x}$. The other parameters are provided in the table \ref{table}.}
\label{FHOHe}
\end{figure}

Figure \ref{FHOHe} compares the SLD quantum Fisher information with the classical Fisher information obtained from homodyne and heterodyne detection schemes for the estimation of $g$ and $k_x$. It is clearly observed that the SLD-based QFI provides the largest estimation sensitivity, representing the ultimate quantum precision limit achievable in the system. Among the experimentally realistic measurement strategies, heterodyne detection yields larger CFI values than homodyne detection over the entire evolution time. This behavior originates from the fact that heterodyne detection simultaneously measures both quadratures of the output field, thereby extracting a larger amount of information encoded in the Gaussian fluctuations. In contrast, homodyne detection only probes a single quadrature component, limiting the accessible parameter information. Furthermore, the decay of all Fisher information curves with time reflects the detrimental role of decoherence and dissipation on the estimation process. These results demonstrate that heterodyne detection constitutes the most efficient experimentally feasible measurement protocol for multiparameter quantum estimation in the considered exciton-optomechanical system.

\section{\label{sec7} CONCLUSION}

In this work, we have investigated multiparameter quantum estimation in a hybrid exciton--optomechanical system by exploiting the Gaussian-state formalism together with the symmetric logarithmic derivative (SLD) and right logarithmic derivative (RLD) approaches. Focusing on the simultaneous estimation of the exciton--photon coupling strength $g$ and the excitonic decay rate $k_x$, we derived the corresponding quantum Fisher information matrices and quantum Cram\'er--Rao bounds, thereby establishing the ultimate precision limits attainable within the considered hybrid platform. Our results reveal that the metrological performance is governed by a subtle interplay between coherent exciton--photon--phonon hybridization, dissipation processes, and thermal fluctuations. In particular, strong hybrid interactions and low-temperature operation substantially enhance the estimation sensitivity, whereas thermal noise and losses progressively deteriorate the attainable precision. We further demonstrated that the comparison between the SLD and RLD bounds provides valuable insight into the multiparameter nature of the estimation problem and allows the identification of the most informative precision limit. In addition, we investigated experimentally accessible Gaussian measurement strategies based on homodyne and heterodyne detection. We found that heterodyne detection generally outperforms homodyne schemes and can approach the ultimate quantum precision limit in suitable parameter regimes. This result highlights the practical feasibility of implementing high-precision multiparameter estimation protocols using currently available continuous-variable measurement techniques. Overall, our findings establish exciton--optomechanical systems as versatile and promising platforms for quantum-enhanced multiparameter metrology. Beyond the specific estimation task considered here, the present work demonstrates how hybrid semiconductor architectures can exploit coherent light--matter interactions and Gaussian quantum fluctuations to achieve enhanced sensing capabilities.

\end{document}